\documentclass[aps,prl,twocolumn]{revtex4}
\usepackage{epsfig,amssymb}
\usepackage[10pt]{moresize}
\begin{document}
\title{Experimental demonstration of a BDCZ quantum repeater node}

\author{Zhen-Sheng Yuan$^{1,2*}$,
Yu-Ao Chen$^{1,2*}$, Bo Zhao$^1$, Shuai Chen$^1$, J\"{o}rg
Schmiedmayer$^{3}$ \& Jian-Wei Pan$^{1,2}$}

\address{$^1$Physikalisches Institut, Ruprecht-Karls-Universit\"{a}t
Heidelberg,
Philosophenweg 12, 69120 Heidelberg, Germany \\
$^2$Hefei National Laboratory for Physical Sciences at Microscale
and Department of Modern Physics, University of Science and
Technology
of China, Hefei, Anhui 230026, China \\
$^3$Atominstitut der \"{O}sterreichischen Universit\"{a}ten,
TU-Wien, A-1020 Vienna Austria\\
$^*$These authors contributed equally to this work. }

\maketitle

\textbf{Quantum communication is a method that offers efficient and secure ways for the exchange of
information in a network. Large-scale quantum communication \cite{PengPRL2007, RosenbergPRL2007,
SchmittPRL2007, UrsinNphys2007} (of the order of 100 km) has been achieved; however, serious
problems occur beyond this distance scale, mainly due to inevitable photon loss in the transmission
channel. Quantum communication eventually fails \cite{GisinRMP2002} when the probability of a dark
count in the photon detectors becomes comparable to the probability that a photon is correctly
detected. To overcome this problem, Briegel, D\"{u}r, Cirac and Zoller (BDCZ) introduced the
concept of quantum repeaters \cite{BriegelPRL1998}, combining entanglement swapping
\cite{ZukowskiPRL1993} and quantum memory to efficiently extend the achievable distances. Although
entanglement swapping has been experimentally demonstrated \cite{PanPRL1998}, the implementation of
BDCZ quantum repeaters has proved challenging owing to the difficulty of integrating a quantum
memory. Here we realize entanglement swapping with storage and retrieval of light, a building block
of the BDCZ quantum repeater. We follow a scheme \cite{ZhaoPRL2007, ChenzbPRA2007} that
incorporates the strategy of BDCZ with atomic quantum memories \cite{DuanNature2001}. Two atomic
ensembles, each originally entangled with a single emitted photon, are projected into an entangled
state by performing a joint Bell state measurement on the two single photons after they have passed
through a 300-m fibre-based communication channel. The entanglement is stored in the atomic
ensembles and later verified by converting the atomic excitations into photons. Our method is
intrinsically phase insensitive and establishes the essential element needed to realize quantum
repeaters with stationary atomic qubits as quantum memories and flying photonic qubits as quantum
messengers.}

Although the BDCZ protocol \cite{BriegelPRL1998} attracted much
interest as a solution to extend the communication length, the
absence of quantum memory has hindered the implementation of quantum
repeaters. In 2001, Duan, Lukin, Cirac and Zoller (DLCZ) proposed an
alternative quantum repeater scheme \cite{DuanNature2001} where
linear optics and atomic ensembles are used to incorporate
entanglement connection and quantum memory into a single unit.
Motivated by the DLCZ protocol, number-state entanglement between
two atomic ensembles has been observed \cite{MatsukevichSCI2004,
ChouNature2005}. Most recently, a functional quantum node
\cite{ChouScience2007} based on asynchronous preparation of
number-state entanglement for two pairs of atomic ensembles -- the
basic element of the DLCZ protocol -- has also been demonstrated.

However, two serious drawbacks make the DLCZ-type functional quantum nodes \cite{DuanNature2001,
ChouScience2007} unlikely to be a realistic solution for long-distance quantum communication
\cite{ZhaoPRL2007, ChenzbPRA2007, JiangPRA2007}. First, the required long-term sub-wavelength
stability of the path difference between two arms of a large scale single-photon interferometer
spanning the whole communication distance is very difficult to achieve \cite{ZhaoPRL2007,
ChenzbPRA2007}, even with the latest and most sophisticated technology for coherent optical phase
transfer \cite{ForemanPRL2007}. Second, the swapping of number-state entanglement using a
single-photon interferometer leads to the growth of a vacuum component in the generated state, and
to the rapid growth of errors due to multiple emissions from individual ensembles
\cite{JiangPRA2007}.

A novel solution \cite{ZhaoPRL2007, ChenzbPRA2007}, is to combine the atomic quantum memory in DLCZ
and the strategy of BDCZ. Since in this scheme two-photon interference is used to generate
long-distance entanglement, the stability requirement for the path differences is determined by the
coherence length of the photons and is consequently 7 orders of magnitude looser \cite{YuanPRL2007}
than in the DLCZ protocol. Moreover, the vacuum component can be suppressed and is no longer a
dominant term after a few entanglement connections \cite{ZhaoPRL2007, ChenzbPRA2007}. Following
this scheme, we demonstrated the implementation of a quantum repeater node, involving entanglement
swapping with the function of storage and retrieval of light. A high precision of local operations
has been achieved that surpasses the theoretical threshold \cite{BriegelPRL1998} required for the
realization of robust quantum repeaters for long-distance quantum communication.

\begin{figure*}
\begin{center}
\epsfig{file=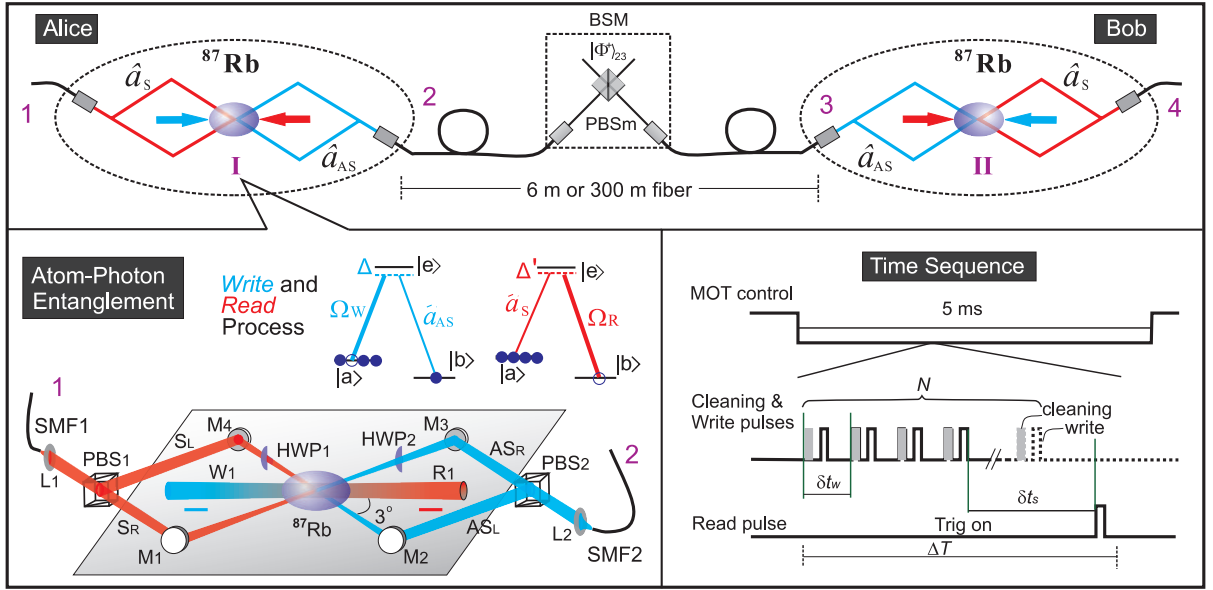,width=16cm}
\end{center}
\caption{The experimental scheme for entanglement swapping. Upper
Panel: photons 2 and 3 overlap at BSM and are projected to the state
$|\Phi^+\rangle_{2,3}$ through which the entanglement is generated
between the two atomic ensembles I and II confined by
magneto-optical traps (MOTs) in different glass cells separated by
$\sim$60 cm. Here 1, 2, 3 and 4 indicate four photons emerging from
the anti-Stokes modes ($\hat{a}_{AS}$) and Stokes modes
($\hat{a}_{S}$), and $|\Phi^+\rangle_{2,3}$ is one of the four Bell
states. Lower-left Panel: atom-photon entanglement. Shown are energy
levels
$\{|a\rangle,~|b\rangle,~|e\rangle\}=\{|5S_{1/2},F=2\rangle,~|5S_{1/2},F=1\rangle,
~|5P_{1/2},F=2\rangle\}$ and the configuration of light beams. PBS,
polarizing beam splitter; HWP, half-wave plate; L, lens; M, mirror;
SMF, single-mode fibre; W, write beam; R, read beam; S, Stokes
field; AS, anti-Stokes field; $\Omega$, Rabi frequency of light
fields. Lower-right Panel: The time sequence of the experimental
procedure at each site. For 6-m (300-m) fibre connection, there are
250 (200) experiment cycles in 5 ms and $\Delta T$ is 16 $\mu$s (20
$\mu$s) for one cycle, which contains $N$=10 ($N$=8) write
sequences. The interval between two neighbouring write pulses is
$\delta t_w=1~\mu$s (1.5 $\mu$s) and $\delta t_s$ is the storage
time. Whenever there is a desired coincidence event between photons
2 and 3, the following write sequence is stopped by a feedback
circuit and the retrieve process can be started (at the time point
labelled `Trig.on'). } \label{fig:setup}
\end{figure*}

In our experiment, to demonstrate entanglement swapping with storage and retrieval of light, we
follow three steps: implementing two atom-photon entanglement sources, sending the flying qubits
(the photons) to an intermediate station for a BSM, and verifying the entanglement between the
stationary qubits (the two remote atomic ensembles).

Unlike previous atom-photon entanglement sources realized with trapped ions
\cite{MoehringNature2007}, single atoms in a cavity \cite{WilkScience2007}, or two spatially
separated atomic ensembles \cite{ChenNPhys2007}, we use two collective excitations in different
spatial modes of a single atomic ensemble to implement the atom-photon entanglement
\cite{ChenShuaiPRL2007}. In contrast to the method in which two separated spatial regions in one
atomic cloud are covered by their own read and write beams \cite{ChouScience2007}, here the two
excitation modes share the same write and read beams, which offers high-quality entanglement and
long-term stability.

The basic principle is shown in Fig. 1 (see Methods). Alice and Bob each have a cold atomic
ensemble consisting of about $10^8$ $^{87}$Rb loaded by magneto-optical traps (MOTs). At each site
atoms are first prepared in the initial state $|a\rangle$, followed by a weak \textit{write} pulse.
Two anti-Stokes fields $AS_L$ and $AS_R$ induced by the write beam via spontaneous Raman scattering
are collected at $\pm 3^\circ$ relative to the propagating direction of the write beam. This
defines two spatial modes of excitation in the atomic ensembles (\textit{L} and \textit{R}), which
constitute our memory qubit.

The two anti-Stokes fields in modes \textit{L} and \textit{R} are
adjusted to have equal excitation probability and orthogonal
polarizations. The two fields are then overlapped at a polarizing
beam splitter PBS2 and coupled into a single-mode fiber. Neglecting
the vacuum state and higher order excitations, the entangled state
between the atomic and photonic qubits can be described effectively
as,
\begin{equation}\label{eqn:entangle1}
|\Psi\rangle_{\mbox{\ssmall
at-ph}}=\frac{1}{\sqrt{2}}\left(|H\rangle|R\rangle+e^{i\phi_{1}}|V\rangle|L\rangle\right)
\end{equation}
where $|H\rangle$/$|V\rangle$ denotes horizontal/vertical
polarization of the single anti-Stokes photon and
$|L\rangle$/$|R\rangle$ denotes single collective excitation in
ensemble $L$/$R$, $\phi_{1}$ is the propagating phase difference
between the two anti-Stokes fields before they overlap at PBS2.
Physically, the atom-photon entangled state (\ref{eqn:entangle1}) is
equivalent to the maximally polarization-entangled state generated
by spontaneous parametric down-conversion \cite{KwiatPRL1995}.

In this way, one can implement two separate and remote atom-photon entanglement sources at Alice
(I) and Bob's (II) sites respectively. To make the higher order excitations negligible, a low
excitation probability ($\chi_m\sim0.01$) is chosen. Due to the imperfect coupling of light modes,
the transmission loss, and the inefficiency of single photon detectors, the overall detection
efficiency of an emerging anti-Stokes photon ($\eta_{\mbox{\ssmall AS}}$) is around $25\%$. To
check the quality of atom-photon entanglement, a read pulse (see Methods) is applied after a
controllable time-delay $\delta t_s$ to convert the atomic collective excitation back into a Stokes
field. Ideally, the retrieve efficiency of the Stokes fields should reach unity. However, various
imperfections such as low optical depth of the atomic ensembles and mode mismatching between the
write and read pulses lead to a 35\% retrieve efficiency. Together with the non-ideal collection
and detection efficiency ($\sim$40\%) of single photon detectors, the overall detection efficiency
of the Stokes photon is around 15\%. After combining the two retrieved Stokes fields on PBS1 (see
Fig. 1), the anti-Stokes and Stokes fields are in the following maximally polarization-entangled
state
\begin{equation}\label{eqn:stokesAntistokes}
|\Psi\rangle_{\mbox{\ssmall
AS,S}}=\frac{1}{\sqrt{2}}\left(|H\rangle_{\mbox{\ssmall
AS}}|H\rangle_{\mbox{\ssmall S}} +
e^{i(\phi_{1}+\phi_{2})}|V\rangle_{\mbox{\ssmall
AS}}|V\rangle_{\mbox{\ssmall S}}\right),
\end{equation}
where $\phi_{2}$ represents the propagating phase difference between two Stokes fields before they
overlap at PBS1. In our experiment, the total phase $\phi_{1}+\phi_{2}$ is actively stabilized via
the built-in Mach-Zehnder interferometer and fixed to zero (see the online supplementary
information). With a time-delay $\delta t_s=$1 $\mathrm{\mu s}$, the measured polarization
correlations of the Stokes and anti-Stokes photons show a strong violation of a CHSH-type Bell's
inequality, with a visibility of $92\%$, confirming the high quality of our atom-photon
entanglement sources. Further measurement shows our atom-photon entanglement still survives up to a
storage time of $\delta t_s=$20 $\mathrm{\mu s}$ (see the online supplementary information).

\begin{figure}
\begin{center}
\epsfig{file=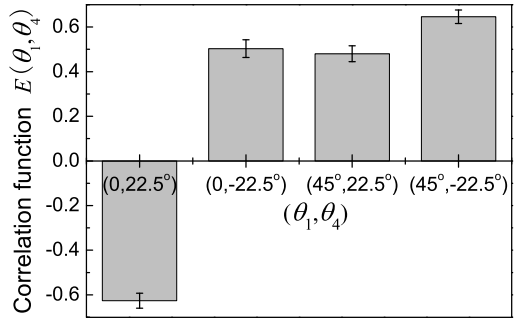, width=8cm}
\end{center}
\caption{Correlation functions of a CHSH-type Bell's inequality with
the storage time $\delta t_s=500$ ns. Error bars represent
statistical errors, which are $\pm1$ standard
deviation.}\label{fig:BellInequality}
\end{figure}

\begin{figure}
\begin{center}
\epsfig{file=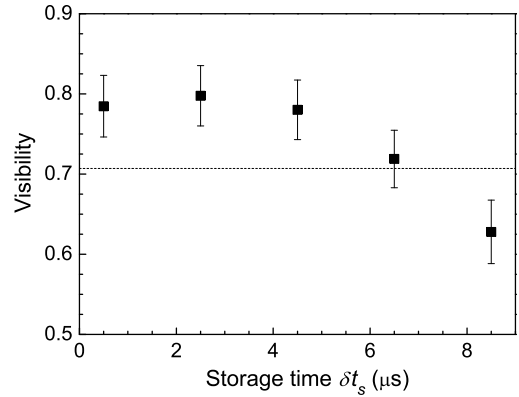,width=8cm}
\end{center}
\caption{Visibility as a function of the storage time with 6 m fiber
connection. Black dots are for the visibility and the dashed line
shows the threshold for the violation of the CHSH-type Bell's
inequality. Error bars represent statistical errors, which are
$\pm1$ standard deviation.}\label{fig:visibility}
\end{figure}

We now describe the entanglement generation between atomic ensembles I and II via entanglement
swapping. As shown in Fig. \ref{fig:setup}, photon 2 from Alice and photon 3 from Bob are both sent
through a 3 m optical fiber to an intermediate station for a joint BSM. In the experiment, we chose
to analyze the projection onto the Bell state
$|\Phi^{+}\rangle_{2,3}=\frac{1}{\sqrt{2}}\left(|H\rangle_2|H\rangle_3
+|V\rangle_2|V\rangle_3\right)$, which is achieved by overlapping photons 2 and 3 onto a polarizing
beam splitter (PBSm) and performing a proper polarization decomposition in the output modes and a
subsequent coincidence detection \cite{PanPRA1998}. Conditioned on detecting a
$|\Phi^{+}\rangle_{2,3}$ state at the intermediate station, the two remote atomic ensembles is
projected onto an identical entangled state $|\phi^{+}\rangle_{\mbox{\ssmall
I,II}}=\frac{1}{\sqrt{2}}\left(|R\rangle_{\mbox{\ssmall I}}|R\rangle_{\mbox{\ssmall
II}}+|L\rangle_{\mbox{\ssmall I}}|L\rangle_{\mbox{\ssmall II}}\right)$
\cite{ZukowskiPRL1993,PanPRL1998}.

It is worth to note that double excitations in either atomic
ensemble I or II will cause false events in the BSM
\cite{ZhaoPRL2007,ChenzbPRA2007}, which reduce the success
probability of entanglement swapping by a factor of 2.
Experimentally, the false events can be eliminated at the stage of
entanglement verification by the four-fold coincidence measurement
of photons 1, 2, 3 and 4. Note that, the detection time of photons 1
and 4 is later than that of photons 2 and 3 by an interval $\delta
t_s$, the storage time in quantum memories. More importantly, such
false events do not affect the applications of our experimental
method in quantum repeaters, since the generation of entanglement
will be deterministic after a second step of connecting two such
nodes, where double excitations are excluded automatically
\cite{ZhaoPRL2007,ChenzbPRA2007}.

The established entanglement between atomic ensembles I and II can
be verified by converting the atomic spins into an entangled photon
pair 1 and 4, which are in the state $|\Phi^{+}\rangle_{1,4}$. Here
we measure the $S$ parameter in a CHSH-type Bell's inequality,
\begin{equation}\label{eqn:chsh}
S=|E(\theta_1,\theta_4)-E(\theta_1,\theta_4^\prime)-
E(\theta_1^\prime,\theta_4)- E(\theta_1^\prime,\theta_4^\prime)|,
\end{equation}
where $E(\theta_1,\theta_4)$ is the correlation function and,
$\theta_1$ and $\theta_1^\prime$ ($\theta_4$ and $\theta_4^\prime$)
are the measured polarization bases of photon 1 (4). In the
measurement, the polarization settings are ($0^\circ$,$22.5^\circ$),
($0^\circ$,$-22.5^\circ$), ($45^\circ$,$22.5^\circ$) and
($45^\circ$,$-22.5^\circ$), respectively.

At a storage time $\delta t_s$=500 ns, the measured correlation functions (shown in Fig.
\ref{fig:BellInequality}) result in $S=2.26\pm0.07$, which violates Bell's inequality by 3 standard
deviations. To observe the lifetime of the entanglement between two remote memory qubits, we
measure the interference visibility of photons 1 and 4 as a function of the storage time by
choosing the polarization basis of $+/-$ (shown in Fig.\ref{fig:visibility}, with
$|+\rangle=\frac{1}{\sqrt{2}}(|H\rangle+|V\rangle)$ and
$|-\rangle=\frac{1}{\sqrt{2}}(|H\rangle-|V\rangle)$ ). Up to a storage time of 4.5 $\mu$s, the
visibility is still higher than the threshold $1/\sqrt{2}$, sufficient for the violation of Bell's
inequality. From the visibilities of the atom-photon and atom-atom entanglements, the precision of
local operations at the BSM station is estimated to be better than 97\% (see the online
supplementary information). We emphasize that this precision achieved here surpasses the threshold
of 95\% for local operations of independent photons necessary for future entanglement purification
and connections \cite{BriegelPRL1998}, and therefore fits the requirement for a scalable quantum
network.

\begin{figure}
\begin{center}
\epsfig{file=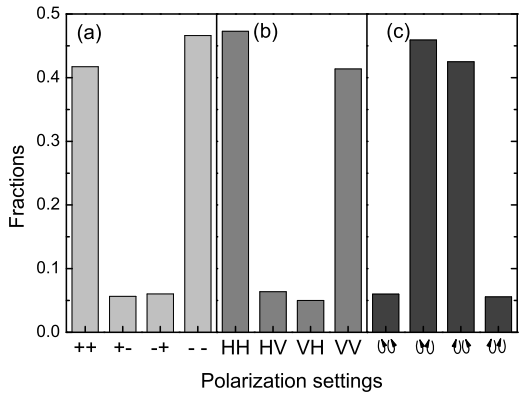,width=8cm}
\end{center}
\caption{Experimental outcomes of the fractions at different
polarization settings with 300 m fiber connection. The polarization
bases are chosen as (a) $|+\rangle$ and $|-\rangle$, (b) $|H\rangle$
and $|V\rangle$, and (c)
\hbox{$|\hspace{-0.1cm}\circlearrowleft\rangle$} and
\hbox{$|\hspace{-0.1cm}\circlearrowright\rangle$}
respectively.}\label{fig:tomo}
\end{figure}

To demonstrate the robustness of our protocol in generation of
quantum entanglement between two atomic ensembles over large
distances, we change the length of the two connecting fibers from 3
m to 150 m. The anti-Stokes photon is delayed 730 ns and the
connection length between Alice and Bob is 300 m. The entanglement
swapping can be quantified by the fidelity of the measured state of
the atomic ensembles. To determine the fidelity, we write the
density matrix of  $|\phi^{+}\rangle_{\mbox{\ssmall I,II}}$ in terms
of the Pauli matrices:
\begin{eqnarray}\label{fidelity}
|\phi^{+}\rangle\langle\phi^{+}|_{\mbox{\ssmall
I,II}}=\frac{1}{4}\left(I+\hat{\sigma}_x\hat{\sigma}_x-\hat{\sigma}_y\hat{\sigma}_y+
\hat{\sigma}_z\hat{\sigma}_z\right).
\end{eqnarray}

Here $\sigma_x=|+\rangle\langle +|- |-\rangle\langle -|$,
$\sigma_y=|\hspace{-0.1cm}\circlearrowright\rangle\langle \circlearrowright|-
|\hspace{-0.1cm}\circlearrowleft\rangle\langle \circlearrowleft|$, and $\sigma_z=|H\rangle\langle
H|- |V\rangle\langle V|$, with
\hbox{$|\hspace{-0.1cm}\circlearrowleft\rangle$}=$\left(1/\sqrt{2}\right)\left(|H\rangle+i|V\rangle\right)$
and
\hbox{$|\hspace{-0.1cm}\circlearrowright\rangle$}=$\left(1/\sqrt{2}\right)\left(|H\rangle-i|V\rangle\right)$
. After a storage time of 1230 ns (with a 730 ns delay being taken into account), the two retrieved
photons 1 and 4 are sent to their own polarization analyzer. Three series of polarization settings
are used and the measured local observables are shown in Fig. 4. The fidelity of final state
$\rho_{\mbox{\ssmall exp}}$ on $|\phi^{+}\rangle_{\mbox{\ssmall I,II}}$  is given by
$F=\mbox{Tr}(\rho_{\mbox{\ssmall exp}}|\phi^{+}\rangle\langle\phi^{+}|_{\mbox{\ssmall
I,II}})=0.83\pm0.02$, with 2.5 standard deviations beyond the threshold of 0.78 to violate the
CHSH-type Bell's inequality for Werner states, demonstrating the success of entanglement swapping.
This fidelity is comparable to the average value achieved in the DLCZ-type functional quantum node
\cite{ChouScience2007}.

In summary, we have successfully demonstrated high precision entanglement swapping with storage and
retrieval of light, a building block for quantum repeaters. The extension of our work to longer
chains will involve many quantum repeater nodes. To achieve this ambitious goal, several
quantities---such as the lifetime and retrieve efficiency of the quantum memory, the fidelity and
generation rate of the entanglement state---still need to be improved significantly. We suggest
three ways forward. First, better compensation of the residual magnetic field and trapping the
atoms in ``clock states'' \cite{HarberPRA2002} with a blue-detuned optical trap \cite{KugaPRL1997}
should improve the lifetime to $\sim$1 s. Second, a high optical density of the atomic cloud,
achieved by the help of traps or by coupling the atoms into an optical cavity \cite{SimonPRL2007},
should increase the retrieve efficiency close to unity. These improvements of the quantum memory
would greatly enhance the fidelity and generation rate of the entanglement. Last, by local
generation of entangled pairs of atomic excitations together with the present technique of
entanglement swapping, the entanglement distribution rate can be greatly improved
\cite{SangouardPRA2008}. Not only does our work enable immediate experimental investigations of
various quantum information protocols, but---with the abovementioned future improvements---
entanglement swapping with storage and retrieval of light would also open the way to long-distance
quantum communication.

\vspace{0.5cm}
\noindent\textbf{Methods}

As shown in Fig. 1, Alice and Bob each have a cold atomic ensemble consisting of about 10$^8$ atoms
of $^{87}$Rb with temperature $\sim$100 $\mu$K. After 20 ms of loading atoms into their MOTs
separated by $\sim$60 cm, we switch off the laser beams and magnetic fields of the MOTs and start a
5-ms-long experiment cycle. At each site, atoms are first prepared in the initial $|a\rangle$,
followed by a (50 ns long,$\sim1\mu$W) weak write pulse, which has a beam waist of 240 $\mu$m and
is 10 MHz red-detuned from the $|a\rangle\rightarrow|e\rangle$ transition. Two anti-Stokes fields
$AS_L$ and $AS_R$ induced by the write beam via spontaneous Raman scattering are collected at
$\pm3^\circ$ relative to the propagating direction of the write beam (70 $\mu$m waist,
$|e\rangle\rightarrow|b\rangle$). The excitation probability ($\chi_m$) of the collective modes $m$
($m=L,~R$) is low ($\chi_m\ll 1$); thus the state of the atom-photon field can be expressed as
\cite{DuanNature2001}
\begin{equation}
|\Psi\rangle_{m}\sim|0_{\mbox{\ssmall AS}}0_b\rangle_{m}+
\sqrt{\chi_m}|1_{\mbox{\ssmall AS}}1_b\rangle_{m}+O(\chi_m),
\end{equation}
and $|i_{\mbox{\ssmall AS}}i_b\rangle_{m}$ denote the $i$-fold
excitation of the anti-Stokes field and the collective spin in the
atomic ensemble. The read beam is counter-propagating and
mode-matched with the write beam with a pulse length of 50 ns, a
power of 60 $\mu$W and a frequency close to resonance of the
$|b\rangle\rightarrow|e\rangle$ transition.

\noindent\textbf{Acknowledgements} We thank Wolfgang D\"{u}r for discussion. This work was
supported by the Deutsche Forschungsgemeinschaft, the Alexander von Humboldt Foundation, and the
European Commission through the Marie Curie Excellence Grant and the ERC Grant. This work was also
supported by the National Fundamental Research Program (Grant No.2006CB921900), the CAS, and the
NNSFC.

\noindent\textbf{Author Information} The authors declare no
competing financial interests. Correspondence and requests for
materials should be addressed to Y.A.C
(yuao@physi.uni-heidelberg.de) and J.W.P
(jian-wei.pan@physi.uni-heidelberg.de).

\noindent\textbf{Online supplementary information}\\
http://www.nature.com/nature/journal/v454/n7208/\\
suppinfo/nature07241.html

\end{document}